\documentclass[%
 reprint,
 superscriptaddress,
 amsmath,amssymb,
 aps,
]{revtex4-2}

\usepackage{graphicx}
\usepackage{dcolumn}
\usepackage{bm}
\usepackage{mathrsfs}
\usepackage{comment}

\begin{document}

\title{Free-energy density functional for Strauss's model of transitive networks}

\author{Diego Escribano}
\affiliation{%
 Grupo Interdisciplinar de Sistemas Complejos (GISC), Departamento de Matem\'aticas, Universidad Carlos III de Madrid, 28911 Legan\'es, Madrid, Spain
}%
\author{Jos\'e A. Cuesta}%
 \email{cuesta@math.uc3m.es}
\affiliation{%
 Grupo Interdisciplinar de Sistemas Complejos (GISC), Departamento de Matem\'aticas, Universidad Carlos III de Madrid, 28911 Legan\'es, Madrid, Spain
}%
\affiliation{%
 Institute for Biocomputation and Physics of Complex Systems (BIFI), University of Zaragoza, 50018 Zaragoza, Spain
}%

\date{\today}

\begin{abstract}
Ensemble models of graphs are one of the most important theoretical tools to study complex networks. Among them, exponential random graphs (ERGs) have proven to be very useful in the analysis of social networks. In this paper we develop a technique, borrowed from the statistical mechanics of lattice gases, to solve Strauss's model of transitive networks. This model was introduced long ago as an ERG ensemble for networks with high clustering and exhibits a first-order phase transition above a critical value of the triangle interaction parameter, where two different kinds of networks with different densities of links (or, alternatively, different clustering) coexist. Compared to previous mean-field approaches, our method describes accurately even small networks and can be extended beyond Strauss's classical model---e.g.~to networks with different types of nodes. This allows us to tackle, for instance, models with node homophily. We provide results for the latter and show that they accurately reproduce the outcome of Monte Carlo simulations.
\end{abstract}

\keywords{Suggested keywords}
\maketitle


\section{\label{sec:intro}Introduction}

Networks are currently one of the most useful theoretical tools of analysis \cite{newman:2018}, one that finds applications in many different fields, such as biology \cite{jeong:2000,pascual:2005,iranzo:2016,pospelov:2019}, sociology \cite{snijders:2011}, economics \cite{goyal:2007,jackson:2010}, or technology \cite{pastor:2001,carvalho:2009}. Modeling using networks is a two-step process. First of all, we need to identify which elements of a system can play the role of nodes, and which connections, interactions, or relations between pairs of them can play the role of links. In the World Wide Web (WWW), these two elements could be the web pages and the hyperlinks; in a social environment, the individuals and their relationships; in the cell, the proteins and their interactions. The result of this first modeling step is a snapshot of the system cast as a network.

But in most cases this network is just an instance, a single realization of a more general set of networks that we could have obtained if we had  modeled other similar systems (another portion of the WWW, another group of people, another cell). Often, the network is literally a snapshot because the system evolves in time, so at different instants we observe different realizations of the network. In general, when it comes to modeling through networks, what we look for is a model whose generic features---whichever they may be---reproduce those of the particular instances that we observe. In other words, we look for an \emph{ensemble} of networks from which our particular network is a typical element---and this is the second, and most important, modeling step.

A network ensemble (or random graph model) is specified by two elements: the set $\mathscr{G}$ of possible realizations of the network, and a probability distribution $P(G)$ on this set ($G\in\mathscr{G})$. It is only natural to write $P(G)$ as
\begin{equation}
    P(G)=\Xi^{-1}e^{-H(G)}, \qquad \Xi=\sum_{G\in\mathscr{G}}e^{-H(G)},
    \label{eq:Gibbs}
\end{equation}
where the analogy to the Gibbs distribution in statistical mechanics justifies referring to $H(G)$ as the Hamiltonian of the ensemble. This sort of Gibbs's ensembles for networks appeared in the early 80s under the name of Exponential Random Graphs (ERGs) in the context of social network modeling \cite{holland:1981,strauss:1986}, and were inspired by previous work on Markov random fields \cite{spitzer:1971,besag:1974}. Hammersley-Clifford's theorem \cite{besag:1974,honerkamp:1998} provides the conditions under which the probability distribution of a Markov random field has the form \eqref{eq:Gibbs} of a Gibbs's ensemble.

The Hamiltonian $H(G)$ of an ERG is specified as \cite{fronczak:2018}
\begin{equation}
    H(G)=-\sum_{\mu=1}^r\lambda_{\mu}\omega_{\mu}(G),
\end{equation}
where $\{\omega_1(G),\dots,\omega_r(G)\}$ is a set of observables on the network (or graph) $G$. For instance, $H(G)=\lambda L(G)$, with $L(G)$ defined as the number of links (edges) of the graph, is the well-known Erd\H{o}s-R\'enyi model \cite{newman:2018,fronczak:2018}. The so-called `conjugate' parameters $\lambda_{\mu}$ are determined by fixing the averages of the observables
\begin{equation}
    \langle\omega_{\mu}\rangle=\sum_{G\in\mathscr{G}}\omega_{\mu}(G)P(G)
    =\frac{\partial}{\partial\lambda_{\mu}}\log\Xi.
\end{equation}

The obvious connection between ERGs and statistical mechanics allows us to obtain these ensembles in a different way---one that sheds light on their meaning. The distribution \eqref{eq:Gibbs} can be obtained by maximizing the entropy functional $S=-\sum_GP(G)\log P(G)$ subject to the constraint that $P(G)$ must have specific values of the averages of a certain set of observables \cite{park:2004a}. According to the Bayesian interpretation \cite{jaynes:2003}, the probability distribution thus obtained is the most agnostic one, given the information we have---namely, the values of the specified averages. In other words, any other distribution having the same averages would incorporate spurious information that we do not know to be true or false for our system. In this sense, it is optimal in that it maximizes our ignorance beyond the data we have.

Simple models involving single-link observables, such as the (directed or undirected) Erd\H{o}s-R\'enyi model or the reciprocity model, have a simple closed-form solution \cite{park:2004a,fronczak:2018}. However, as soon as the observables involve two or more links, the models become more difficult to analyze---but also more interesting. The simplest model in which links interact is the 2-star model \cite{park:2004a,park:2004b}. This model exhibits a first-order phase transition, when the interaction is strong enough, from a low-density to a high-density phase, which can be accurately obtained, in the thermodynamic limit of very large number of nodes, using a mean-field approximation.

But perhaps the equivalent to the Ising model for ERGs is Strauss's model of transitive networks \cite{strauss:1986,park:2005}. This model enhances the clustering of the networks by introducing an interaction associated to triangles. As the 2-star model, Strauss's can be studied in mean-field approximation, and it also exhibits a similar phase transition.

Ever since they were discovered, researchers have been intrigued by the nature of the phase transition in Strauss's model \cite{jonasson:1999,burda:2004,park:2004a,park:2004b,park:2005,chatterjee:2013,tamm:2014,avetisov:2016,yin:2016}, and have explored extensions of it in pursue of ensembles either with more realistic features \cite{holme:2002,newman:2009,foster:2010,bianconi:2014,avetisov:2016,lopez:2021} or more amenable to analytic calculations \cite{lopez:2018,lopez:2020}. What most of these studies seem to imply is that, for certain sets of parameters, there are values of some observables (e.g.~the mean number of links or the clustering) that no graph in the ensemble can attain. Furthermore, in this regime the typical graphs of the ensemble abruptly change from sparse to dense upon a slight variation of the control parameter. These facts cast serious doubts on the usefulness of Strauss's model \cite{park:2005}, or even of ERGs in general \cite{fronczak:2018}, to reproduce the features of real-life networks. For this reason, alternative models have been proposed exhibiting topological and dynamical features more akin to those of real networks \cite{holme:2002,newman:2009,miller:2009,foster:2010,bianconi:2014}---even though they still produce graphs that are hard to tune. More recent models choose to impose a given degree distribution as a control mechanism \cite{tamm:2014,lopez:2018,pospelov:2019,lopez:2020,lopez:2021}, but the phase transition remains and clustered networks are still produced.

The aim of this paper is two-fold. First of all, by introducing the language of lattice gases \cite{palla:2004} we will show that there is no qualitative difference between the phase transition exhibited by Strauss's model and the condensation transition of an Ising lattice gas \cite{lavis:1999}. Thermodynamics teaches us how to interpret states that have intermediate densities between a liquid and a gas. Likewise, thermodynamics will provide a description of the sort of networks that we must expect for those ``forbidden'' values of the observables in Strauss's model.

Secondly, we will address this problem using a density-functional formalism especially tailored for lattice gases \cite{lafuente:2004,lafuente:2005}. This formalism provides a method to construct a mean-field-like free energy of the system, from which everything else can be derived. It also has the advantage that the non-homogeneous counterpart of Strauss's model can be solved with no extra effort. Networks in which nodes of different types interact in different ways are of this kind, and using them we can study, e.g., the effect of homophily in social networks.

The paper is organized as follows. Section~\ref{sec:straussmodel} introduces Strauss's model and its interpretation as a lattice gas of links. The version of the model we will be dealing with is intrinsically inhomogeneous insofar as the interaction parameters are all link-dependent. In Sec.~\ref{sec:FMT} we use a density-functional formalism to obtain the free energy of the system. In Sec.~\ref{sec:uniform} we calculate the free energy assuming that every link has equal probability of occurring. We discuss the thermodynamic limit as well as the well-known phase transition that Strauss's homogeneous model exhibits. The lattice-gas viewpoint we are adopting reveals that this transition is akin to a condensation in fluids, and this analogy allows us to discuss the nature of the system in the region of coexistence. It is one of the main points of this work, because it stands for the validity of the model even in the coexistence region---which has been questioned so far. We end the section by comparing our results to those of Park and Newman \cite{park:2005} for finite networks, where the present approach proves to be more accurate (they both coincide in the thermodynamic limit). Finally, Sec.~\ref{sec:nonhomogeneous} discusses some of the differences that inhomogeneities (e.g., homophily when there are different kinds of nodes) introduce in the system, as a way of illustrating the ability of the functional developed here to deal with this sort of situations. We conclude in Sec.~\ref{sec:discussion} with a summary and a brief discussion.

\section{\label{sec:straussmodel}Strauss's model and its lattice-gas interpretation}

Strauss's model is an ERG ensemble of undirected graphs with $N$ nodes defined by the Hamiltonian
\begin{equation}
    -H(G)=\phi L(G)+\frac{\gamma}{N}T(G),
    \label{eq:Straussclassic}
\end{equation}
where $L(G)$ is the number of links (edges) in the graphs and $T(G)$ the number of triangles (clustering). In this model, a positive $\phi$ enhances the creation of links and a positive $\gamma$ enhances the formation of triangles. (Notice that we are using different sign conventions than those used in previous works \cite{strauss:1986,park:2004a,park:2004b,park:2005,fronczak:2018}, in order to agree with those commonly adopted in the statistical mechanics of lattice gases). The factor $N^{-1}$ in front of $T(G)$ accounts for the fact that there are $O(N^3)$ triangles in the network compared to $O(N^2)$ links. With this factor both terms are comparable for large $N$ if the constants are $O(1)$. The model was introduced by Strauss \cite{strauss:1986} to describe graphs with a clustering higher than that of a typical Erd\H{o}s-R\'enyi graph.

Here we will deal with a non-homogeneous version of this model. In order to write down its Hamiltonian we need to introduce some notation. Let $\mathscr{N}$ denote the set of all nodes of an undirected graph $G$, $\mathscr{N}_2$ the set of all subsets of $2$ elements of $\mathscr{N}$ (hence the potential links, which will be denoted by their indexes $\{i_1,i_2\}$), and $\mathscr{N}_3$ the set of all subsets of $3$ elements of $\mathscr{N}$ (hence the potential triangles, which will be denoted $\{i_1,i_2,i_3\}$). If $|\mathscr{N}|=N$, then $|\mathscr{N}_2|=\binom{N}{2}$, $|\mathscr{N}_3|=\binom{N}{3}$. We also denote $\{\tau_{ij}\}$ the adjacency matrix of the graph $G$, where $\tau_{ij}=1$ if the link $\{i,j\}$ exists in $G$ and $\tau_{ij}=0$ otherwise. Then, in terms of this variables a triangle $\{i,j,k\}$ exists if and only if $\tau_{ij}\tau_{jk}\tau_{ki}=1$. Thus, the Hamiltonian of the non-homogeneous version of Strauss's model can be written as
\begin{equation}
    -H(G)=\sum_{\{ij\}\in\mathscr{N}_2}\phi_{ij}\tau_{ij}
    +\sum_{\{ijk\}\in\mathscr{N}_3}\frac{\gamma_{ijk}}{N}\tau_{ij}\tau_{jk}\tau_{ki}.
\end{equation}
The parameters $\phi_{ij}$ and $\gamma_{ijk}$ are local versions of those of Hamiltonian \eqref{eq:Straussclassic}---hence, a positive $\phi_{ij}$ enhances the creation of the link $\{i,j\}$ and similarly a positive $\gamma_{ijk}$ enhances the formation of the triangle $\{i,j,k\}$.

The variables $\tau_{ij}$ play the role of ``particles'' sitting on the links of the \emph{complete} graph over the set of nodes $\mathscr{N}$. If $\tau_{ij}=1$, it means that the link $\{ij\}$ is occupied by a particle, whereas if $\tau_{ij}=0$, the link is empty. Thus, $G$ can also be interpreted as the configuration of $\binom{N}{2}$ such particles in the complete graph. Under this interpretation $\phi_{ij}$ can be regarded as a (local) chemical potential and therefore $\Xi$ in \eqref{eq:Gibbs} would play the role of the grand partition function of this system \cite{lavis:1999,lafuente:2004,lafuente:2005}.

\begin{figure}
\includegraphics[width=0.45\textwidth]{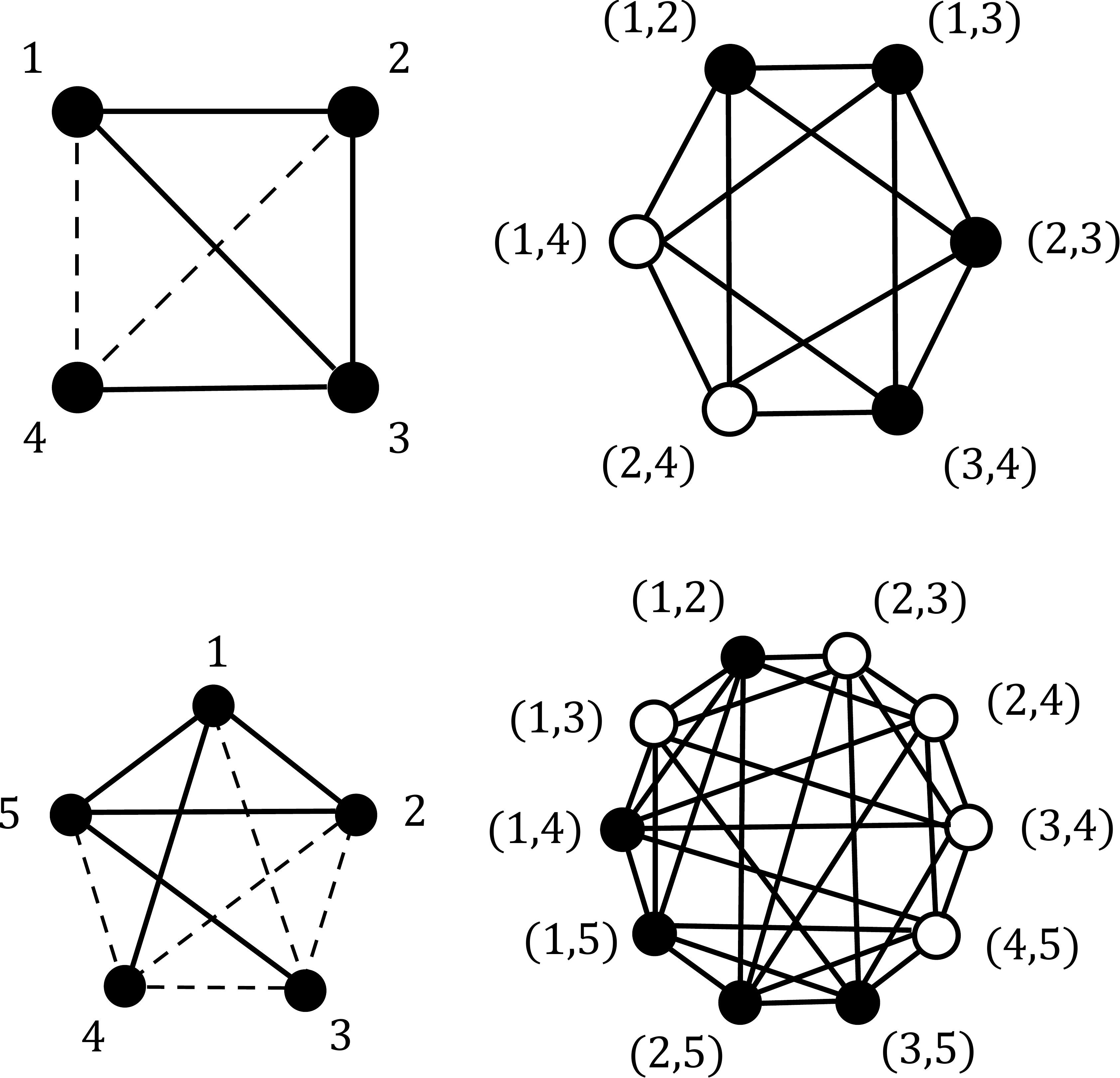}
\caption{\label{fig:dual_networks} Dual networks for the complete graphs of four (up) and five nodes (down). In the dual networks, links are represented as nodes, and two of these nodes are neighbors if they share a node of the original one. (For an alternative representation see~Figure~1 of Ref.~\onlinecite{palla:2004}.)}
\end{figure}

As particles occupy the links, rather than the nodes, of a complete graph, the ``space'' where these particles live in is weird. As a matter of fact, in this dual network every two particles are either neighbors or second neighbors to each other. The reason is that if links $\{ij\}$ and $\{kl\}$ are not neighbors (i.e., have no common nodes) then they are both neighbors to a common link (e.g. $\{ik\}$). Figure~\ref{fig:dual_networks} illustrates these duals networks for the complete graphs of $4$ and $5$ nodes. Each link is neighbor to $2(N-2)$ other links, and second neighbor to the remaining $\binom{N-2}{2}$.

The grand potential of this system $\Omega=-\log\Xi$ is a function of all conjugate fields $\bm{\phi}=\{\phi_{ij}\}$ and $\bm{\gamma}=\{\gamma_{ijk}\}$, from which the probability that link $\{ij\}$ is occupied (henceforth \emph{density}) can be obtained as
\begin{equation}
    -\frac{\partial\Omega}{\partial\phi_{ij}}=\langle\tau_{ij}\rangle=\rho_{ij}.
    \label{eq:densities}
\end{equation}
A Legendre transform on the grand potential yields the free energy
\begin{equation}
    F(\bm{\rho},\bm{\gamma})=\sum_{\{ij\}}\phi_{ij}(\bm{\rho})\rho_{ij}
    +\Omega(\bm{\phi},\bm{\gamma}),
    \label{eq:Legendre}
\end{equation}
where $\phi_{ij}(\bm{\rho})$ is obtained by solving \eqref{eq:densities} for fixed $\bm{\rho}$. Differentiating $F$ with respect to the densities,
\begin{equation*}
    \frac{\partial F}{\partial\rho_{ij}}=\phi_{ij}+\sum_{\{kl\}}
    \frac{\partial\phi_{kl}}{\partial\rho_{ij}}\rho_{kl}
    +\sum_{\{kl\}}\frac{\partial\Omega}{\partial\phi_{kl}}
    \frac{\partial\phi_{kl}}{\partial\rho_{ij}},
\end{equation*}
so if we take \eqref{eq:densities} into account we finally find
\begin{equation}
    \frac{\partial F}{\partial\rho_{ij}}=\phi_{ij},
    \label{eq:chempot}
\end{equation}
which is the usual equation for the chemical potential.

It can be proven that, given the free-energy density functional $F(\bm\rho)$ of a system, its equilibrium density is the unique density profile that minimizes the functional $\Omega(\bm\rho)\equiv F(\bm\rho)-\bm\phi\cdot\bm\rho$ \cite[][Appendix B]{hansen:2013} (alternatively, it minimizes $F(\bm\rho)$ at constant mean density). Thus, Eq.~\eqref{eq:chempot}, which is dual to \eqref{eq:densities}, is the expression of this variational principle. Hence, its solution provides the values of the densities for a given set of chemical potentials $\bm\phi$.

\section{\label{sec:FMT}Fundamental-measure approximation}

The technique we will use to find an approximation to the free energy of Strauss's model is know in the field of lattice gases as \emph{fundamental-measure theory} \cite{lafuente:2004,lafuente:2005}. It is a reformulation of the well-known cluster variation method \cite{lavis:1999}. The idea is to decompose the system in overlapping clusters and express the free energy as a sum of the free energies of those clusters, controlling for overcounting.

In the case of Strauss's model, the geometry of the Hamiltonian suggests that the simplest possible clusters are triangles. Thus, as a first approximation the free energy is obtained as the sum of the contributions to the free energy of all the triangles within the complete graph. However, in doing so every link participates in $N-2$ triangles, so we need to subtract $N-3$ times the contribution to the free energy of all links. In other words, the fundamental-measure approximation to the free energy of this model will be
\begin{equation}
    \begin{split}
        F(\bm\rho,\bm\gamma)=&\,\sum_{\{ijk\}}\Phi_3(\rho_{ij},\rho_{jk},\rho_{ki},\gamma_{ijk}) \\
        &-(N-3)\sum_{\{ij\}}\Phi_2(\rho_{ij}),
    \end{split}
    \label{eq:FMT}
\end{equation}
where $\Phi_2$ and $\Phi_3$ are the free energies of a single link and a single triangle, respectively.

The expression for $\Phi_2$ is easy to obtain. Denoting $z_{ij}\equiv e^{\phi_{ij}}$, the grand partition function for a single link $\{ij\}$ is simply $\Xi_2=1+z_{ij}$. Thus,
\begin{equation*}
    \rho_{ij}=z_{ij}\frac{\partial}{\partial z_{ij}}\log\Xi_2=\frac{z_{ij}}{1+z_{ij}},
\end{equation*}
from which
\begin{equation*}
    z_{ij}=\frac{\rho_{ij}}{1-\rho_{ij}}, \qquad \Xi_2=\frac{1}{1-\rho_{ij}}.
\end{equation*}
Substituting these expressions in the Legendre transform $\Phi_2=\rho_{ij}\log z_{ij}-\log\Xi_2$, we end up with
\begin{equation}
    \Phi_2(\rho_{ij})=\rho_{ij}\log\rho_{ij}+(1-\rho_{ij})\log(1-\rho_{ij}),
    \label{eq:Phi2}
\end{equation}
which is simply the free energy of an ideal lattice gas.

The calculation of $\Phi_3$ is rather more involved, and is deferred to Appendix~\ref{app:A}. Introducing the shorthand
\begin{equation}
    \zeta_{ijk}\equiv \exp(\gamma_{ijk}/N)-1=\frac{\gamma_{ijk}}{N}+O\left(\frac{1}{N^2}\right),
\end{equation}
its expression turns out to be
\begin{equation}
    \begin{split}
        \Phi_3=&\, \Phi_2(\rho_{ij})+\Phi_2(\rho_{jk})+\Phi_2(\rho_{ki})
        +\rho_{ij}\log\left(1-\frac{\rho_{ijk}}{\rho_{ij}}\right) \\
        &+\rho_{jk}\log\left(1-\frac{\rho_{ijk}}{\rho_{jk}}\right)
        +\rho_{ki}\log\left(1-\frac{\rho_{ijk}}{\rho_{ki}}\right) \\
        &-2\log(1-\rho_{ijk}),
    \end{split}
    \label{eq:Phi3}
\end{equation}
where $\rho_{ijk}$ is one of the real solutions of the cubic equation
\begin{equation}
    \zeta_{ijk}(\rho_{ij}-\rho_{ijk})(\rho_{jk}-\rho_{ijk})(\rho_{ki}-\rho_{ijk})=
    \rho_{ijk}\left(1-\rho_{ijk}\right)^2.
    \label{eq:cubic}
\end{equation}
This ``triangle'' density is related to $T_{ijk}=\langle\tau_{ij}\tau_{jk}\tau_{ki}\rangle$, the probability that nodes $i,j,k$ form a triangle, as (see Appendix~\ref{app:B})
\begin{equation}
T_{ijk}=\frac{1+\zeta_{ijk}}{\zeta_{ijk}}\rho_{ijk}
=\frac{1}{1-e^{-\gamma_{ijk}/N}}\rho_{ijk}.
\label{eq:triangleFMT}
\end{equation}

If we now substitute \eqref{eq:Phi2} and \eqref{eq:Phi3} into \eqref{eq:FMT} and take into account that
\begin{equation}
    \sum_{\{ijk\}}(A_{ij}+A_{jk}+A_{ki})=(N-2)\sum_{\{ij\}}A_{ij}
\end{equation}
for any link-dependent magnitude $A_{ij}$, we finally get
\begin{equation}
    \begin{split}
        F
        =&\,\sum_{\{ij\}}\big[\rho_{ij}\log\rho_{ij}+
        (1-\rho_{ij})\log(1-\rho_{ij})\big] \\
        &+\sum_{\{ijk\}}\bigg[\rho_{ij}\log\left(1-\frac{\rho_{ijk}}{\rho_{ij}}\right)
        +\rho_{jk}\log\left(1-\frac{\rho_{ijk}}{\rho_{jk}}\right) \\
        &+\rho_{ki}\log\left(1-\frac{\rho_{ijk}}{\rho_{ki}}\right)-2\log(1-\rho_{ijk})\bigg].
    \end{split}
\label{eq:Ffunctional}
\end{equation}

\section{\label{sec:uniform}Homogeneous networks}

\subsection{\label{sec:freeuniform}Free energy}

We can recover Strauss's original model by assuming $\rho_{ij}=\rho$ for every link $\{ij\}$ and $\gamma_{ijk}=\gamma$ for every triangle $\{ijk\}$. Then the free energy \emph{per link} $f(\rho,\gamma)=\binom{N}{2}^{-1}F(\bm\rho,\bm\gamma)$ will be
\begin{equation}
\begin{split}
f
=&\,\rho\log\rho+(1-\rho)\log(1-\rho) \\
&+(N-2)\left[\rho\log\left(1-\frac{\rho_{\text{T}}}{\rho}\right)
-\frac{2}{3}\log(1-\rho_{\text{T}})\right],
\end{split}
\label{eq:funif}
\end{equation}
where $\rho_{\text{T}}$ is the only real root of
\begin{equation}
\zeta(\rho-\rho_{\text{T}})^3=\rho_{\text{T}}(1-\rho_{\text{T}})^2.
\label{eq:rhoTrho}
\end{equation}
With the change of variable
\begin{equation}
    t=\frac{\rho-\rho_T}{1-\rho_T}, \qquad \rho_T=\frac{\rho-t}{1-t},
\end{equation}
the cubic equation \eqref{eq:rhoTrho} can be rewritten as
\begin{equation}
    t^3+\frac{t}{\zeta(1-\rho)}-\frac{\rho}{\zeta(1-\rho)}=0.
\end{equation}
This equation has only one real root, which is given by the formula \cite[][pp.~102--103]{birkhoff:1997}
\begin{equation}
t=\frac{2}{\sqrt{3\zeta(1-\rho)}}\sinh\left[\frac{1}{3}
\sinh^{-1}\left(\frac{3}{2}\rho\sqrt{3\zeta(1-\rho)}\right)\right].
\label{eq:t}
\end{equation}

\subsection{\label{sec:thlimit}Thermodynamic limit}

In the thermodynamic limit $N\to\infty$ one can see, either from \eqref{eq:rhoTrho} or directly from \eqref{eq:t}, that the (thermodynamic) free energy becomes simply
\begin{equation}
    f_{\text{th}}=\rho\log\rho+(1-\rho)\log(1-\rho)-\frac{\gamma\rho^3}{3}.
    \label{eq:freenergythermo}
\end{equation}
This free energy is convex as long as
\begin{equation}
    \frac{\partial^2f_{\text{th}}}{\partial\rho^2}=\frac{1}{\rho(1-\rho)}-2\gamma\rho>0,
    \label{eq:ddfpos}
\end{equation}
which is equivalent to $2\gamma\rho^2(1-\rho)<1$. As the maximum value of $\rho^2(1-\rho)$ is $4/27$ (reached at $\rho_c=2/3$), the condition above implies $\gamma<27/8$. So, the values $\gamma_c=27/8$, $\rho_c=2/3$, mark a critical point, above which the system exhibits a first order phase transition.

\begin{figure}
\includegraphics[width=0.5\textwidth]{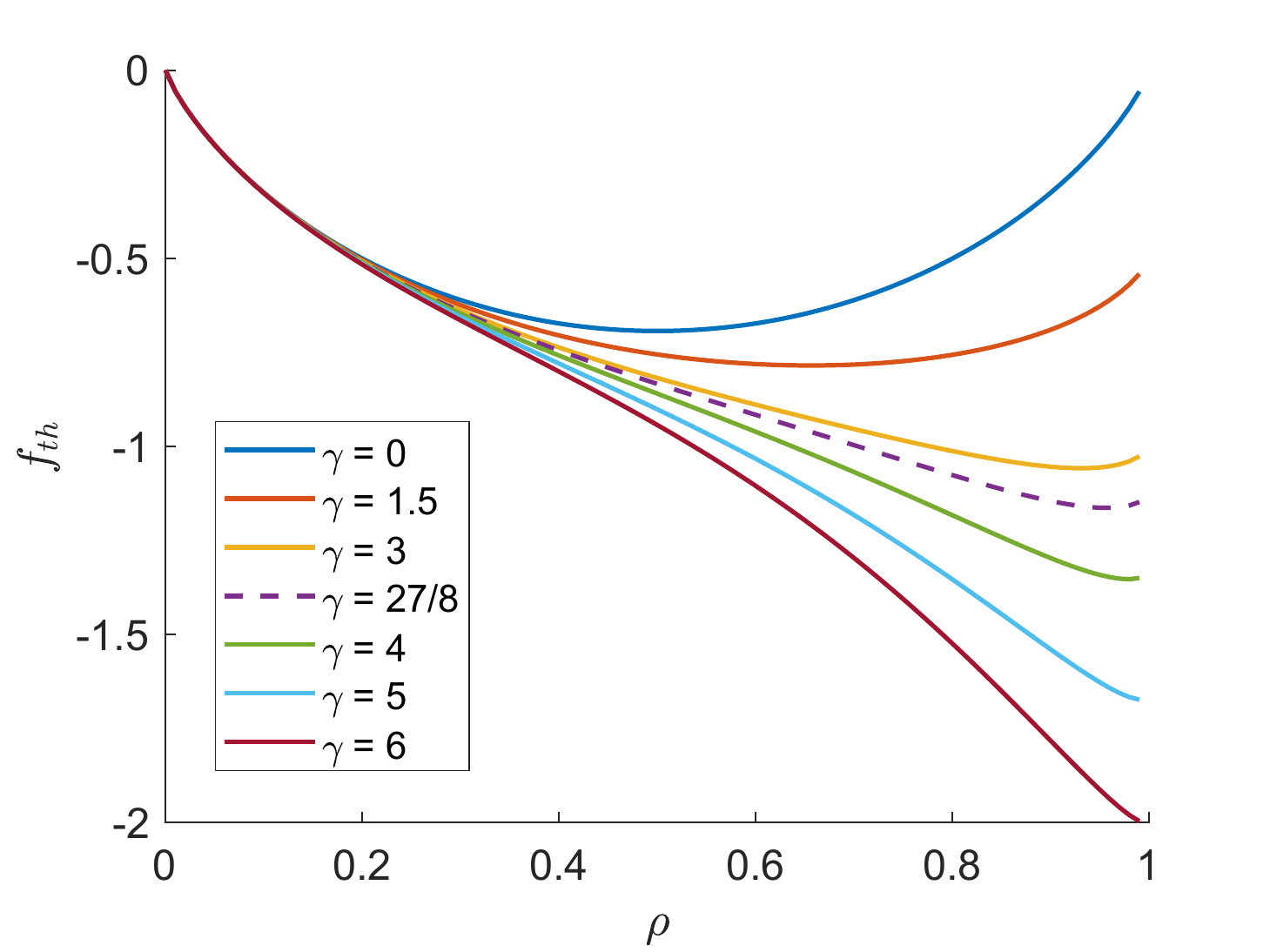}
\caption{\label{fig:fgamma} Thermodynamic free energy of the model for different values of $\gamma$ below and above the critical value $\gamma_c$ ($\gamma$ increases from top to bottom). The curves illustrate the onset of the concavity as $\gamma$ grows past $\gamma_c = 27/8$, which is represented by the purple dashed line.}
\end{figure}

Figure~\ref{fig:fgamma} illustrates the concavity that $f_{\text{th}}$ develops as $\gamma$ increases past $\gamma_c$. It is the fingerprint of a condensation transition in lattice gases \cite{lavis:1999} because a concave free energy implies thermodynamic instability (the compressibility is negative). The homogeneous ``fluid'' separates in two phases, each of a different density, in thermodynamic equilibrium. The fraction occupied by each phase must be such that the overall density matches the prescribed one.

\begin{figure}
\includegraphics[width=0.5\textwidth]{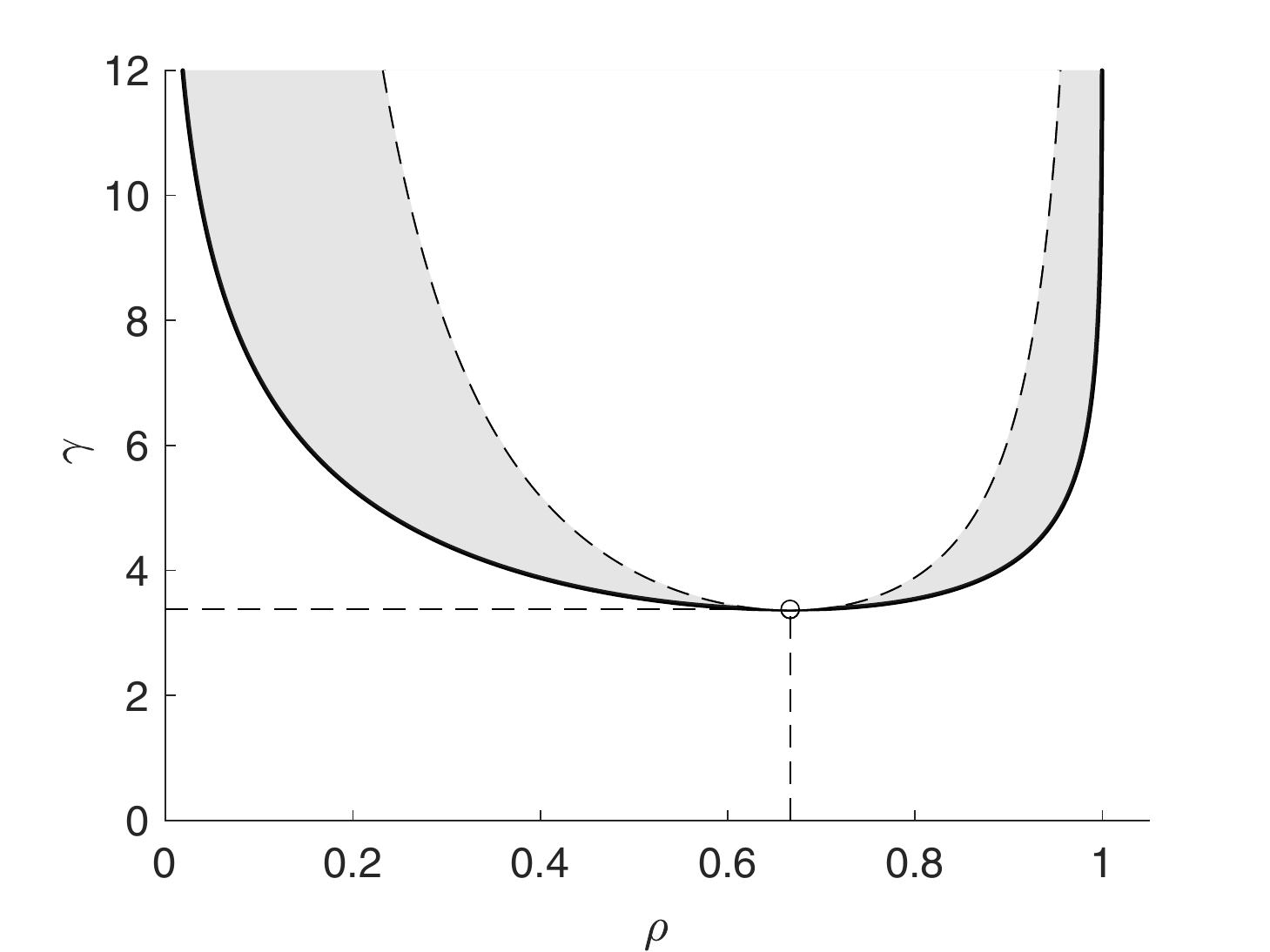}
\caption{\label{fig:coexistence} In the upper region delimited by the solid curve the system is not homogeneous, but separated in two coexisting phases whose respective densities are given by the values of curve at the corresponding $\gamma$. The dashed line represents the spinodal \eqref{eq:spinodal}, i.e.~the curve at which the compressibility vanishes---hence the thermodynamic free energy changes from convex to concave. The circle, where both curves meet, marks the critical point. Within the shaded region the system may be trapped in a metastable, homogeneous state.}
\end{figure}

\begin{figure*}
\includegraphics[width=0.85\textwidth]{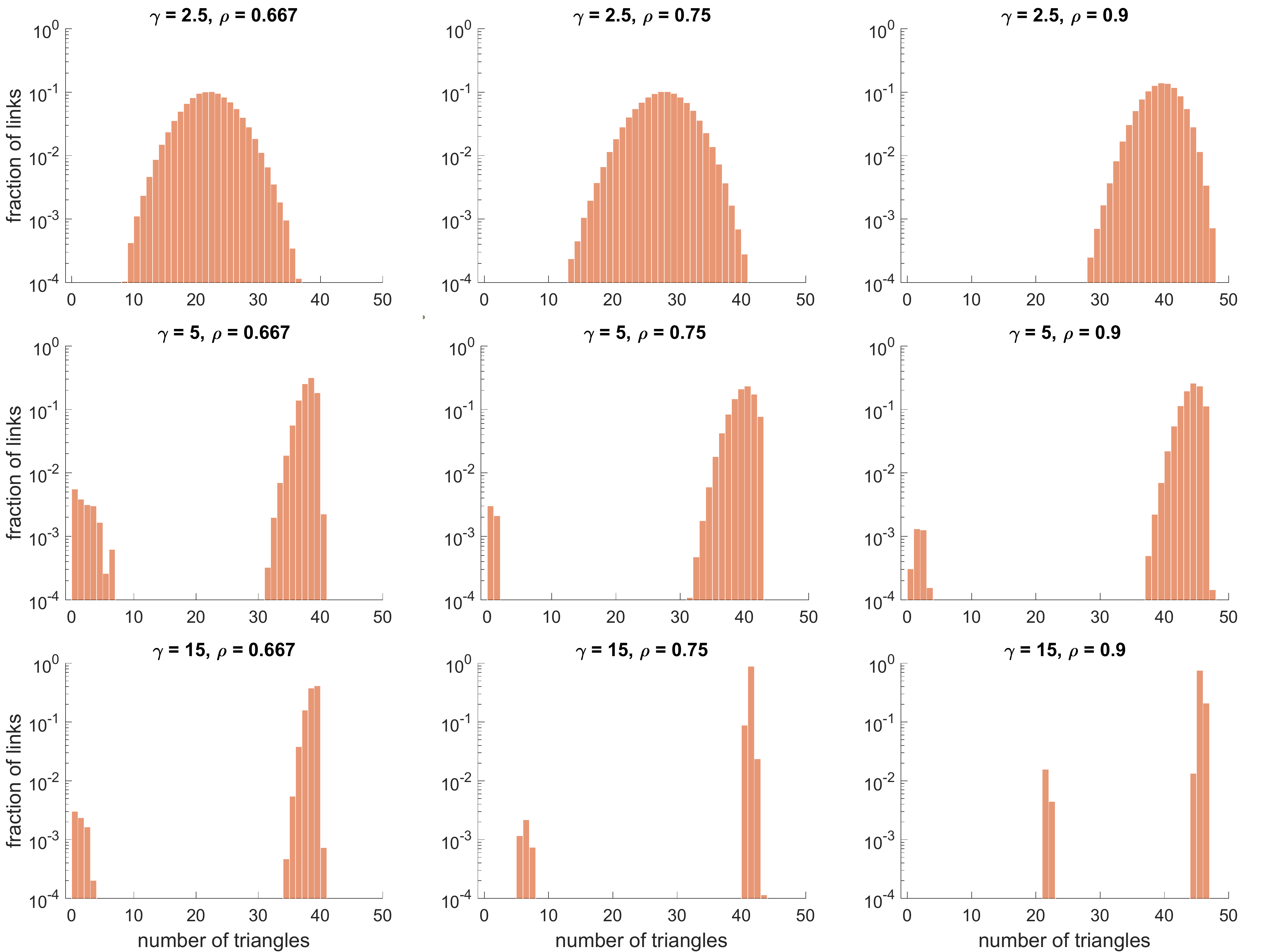}
\caption{\label{fig:histograms} Each of these nine panels depicts the average fraction of links belonging to a given number of triangles, as obtained from Monte Carlo simulations---using Kawasaki dynamics---of a Strauss network with $N=50$ nodes, for three values of the interaction parameter $\gamma$ (below, just above, and well above the critical point) and three different densities.}
\end{figure*}

Thermodynamic equilibrium means ``chemical'' equilibrium (equality of chemical potentials) and ``mechanical'' equilibrium (equality of pressures). The first condition implies $f_{\rho}(\rho_1,\gamma)=f_{\rho}(\rho_2,\gamma)$; the second condition implies $\rho_1f_{\rho}(\rho_1,\gamma)-f(\rho_1,\gamma)=\rho_2f_{\rho}(\rho_2,\gamma)-f(\rho_2,\gamma)$. Both conditions are summarized in the equation
\begin{equation}
    f_{\rho}(\rho_1,\gamma)=f_{\rho}(\rho_2,\gamma)=\frac{f(\rho_2,\gamma)-f(\rho_1,\gamma)}{\rho_2-\rho_1},
\end{equation}
which represent Maxwell's double tangent construction \cite{huang:1987}. For $f_{\text{th}}$, the solution of these equations is represented in Fig.~\ref{fig:coexistence}. Given any $\rho_1<\rho<\rho_2$, there will be a fraction $x$ of the graph of density $\rho_1$ and a fraction $1-x$ of density $\rho_2$ such that $\rho=x\rho_1+(1-x)\rho_2$.

On the other hand, the condition $f_{\rho\rho}(\rho,\gamma)=0$ marks the points where the compressibility vanishes---i.e., where the system is no longer mechanically stable. This curve is known as the \emph{spinodal} (see Fig.~\ref{fig:coexistence}). According to \eqref{eq:ddfpos}, this curve is
\begin{equation}
    \gamma=\frac{1}{2\rho^2(1-\rho)}.
    \label{eq:spinodal}
\end{equation}
Within the region between the coexistence curve and the spinodal (the shaded area of Fig.~\ref{fig:coexistence}) the system can still be prepared in a homogeneous---but metastable---state. This explains the origin of the hysteresis usually observed in first-order phase transitions---this one in particular \cite{park:2005}.

It is difficult to guess the nature of the phase transition that this system undergoes above the critical point. Recall that any two links are separated by no more than one intermediate neighbor. The very notion of ``space'' breaks down in such a system, so the picture of the usual condensation transition, where gas and liquid occupy different portions of the volume, has no reasonable counterpart in a complete graph. Nonetheless, the transition may be illustrated by computing a histogram of the number of links belonging to a given number of triangles \cite{tamm:2014}. As a way of illustration we have obtained such histograms by performing Monte Carlo simulations using the dynamics of Kawasaki \cite{kawasaki:1972}, which preserves the number of links---hence the density $\rho$. In this dynamics, a Monte Carlo step amounts to first removing a link at random and then creating a link also at random. The results, obtained for three different values of the interaction $\gamma$ (below, just above, and well above the critical point) and three different densities, are depicted in Fig.~\ref{fig:histograms}. In each of these simulations we perform $5\times10^5$ Monte Carlo steps. When $\gamma<\gamma_c$ the histograms show a single peak that shifts to the right and shrinks as the density increases, whereas if $\gamma>\gamma_c$ the distribution exhibits two very neat peaks, one at high values and the other one at lower values of the number of triangles. Obviously, the links forming each of the two peaks belong to each of the two---low and high density---phases. Figure~\ref{fig:histograms} reveals that networks within the coexisting region do exist, but they have different structural properties than those outside this region. We will return to this point in Sec.~\ref{sec:discussion}.

\subsection{Finite networks}
We can use an asymptotic expansion in $N$ to obtain $\rho_T$ from \eqref{eq:t}. The first two terms are
\begin{equation}
    \rho_T=\frac{\gamma\rho^3}{N}\left[1+\frac{\gamma(1-6\rho^2+4\rho^3)}{2N}
    +O\left(\frac{1}{N^2}\right)\right],
\end{equation}
and consequently, the free energy can be expanded as
\begin{equation}
    \begin{split}
        f=&\, \rho\log\rho+(1-\rho)\log(1-\rho)-\frac{\gamma\rho^3}{3}\bigg[1-\frac{2}{N} \\
        &+\frac{\gamma(1-3\rho^2+2\rho^3)}{2N}+O\left(\frac{1}{N^2}\right)\bigg].
    \end{split}
    \label{eq:fN}
\end{equation}
The critical point will then be a solution of the equations $f_{\rho\rho}=f_{\rho\rho\rho}=0$, which yields
\begin{equation}
    \begin{split}
        &\gamma_c(N)=\frac{27}{8}\left[1+\frac{45}{16N}+O\left(\frac{1}{N^2}\right)\right], \\
        &\rho_c(N)=\frac{2}{3}+O\left(\frac{1}{N^2}\right).
    \end{split}
    \label{eq:gammacasymp}
\end{equation}

\begin{figure}
\includegraphics[width=0.45\textwidth]{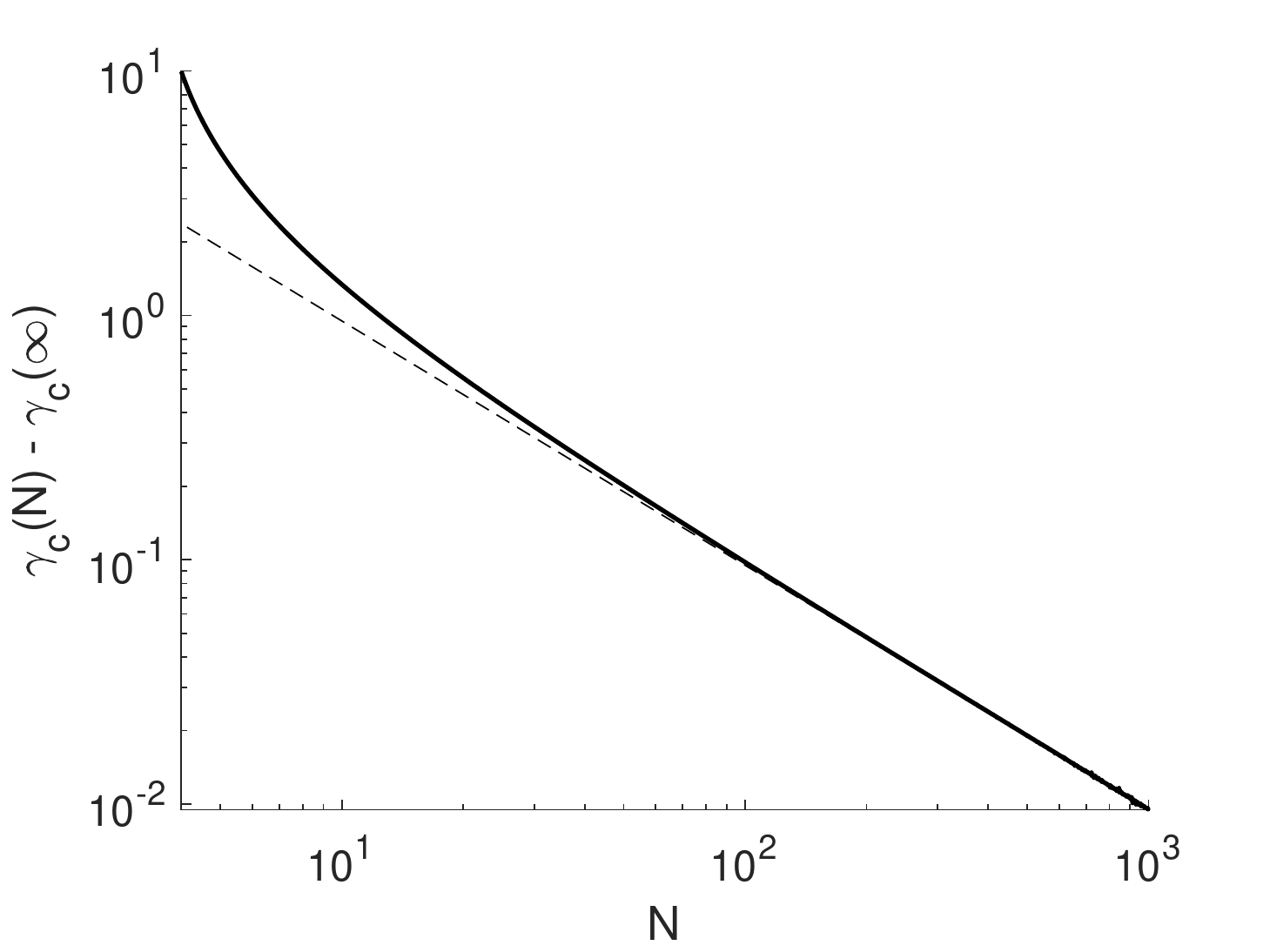}
\caption{\label{fig:gammacN} Difference between the critical value $\gamma_c(N)$ for a network with $N$ nodes and its limit for $N\to\infty$, as a function of $N$. The solid line is obtained by numerically solving the equations for the critical point; the dashed line arises from the asymptotic expression \eqref{eq:gammacasymp}.}
\end{figure}

The numerical solution for $\gamma_c(N)$ is depicted in Fig.~\ref{fig:gammacN} along with the asymptotic expansion above. As for $\rho_c(N)$, within numerical resolution we find that its value is always $2/3$. It is noteworthy that the curve of $\gamma_c(N)$ diverges somewhere between $N=4$ and $N=3$.

\begin{figure}
\includegraphics[width=0.5\textwidth]{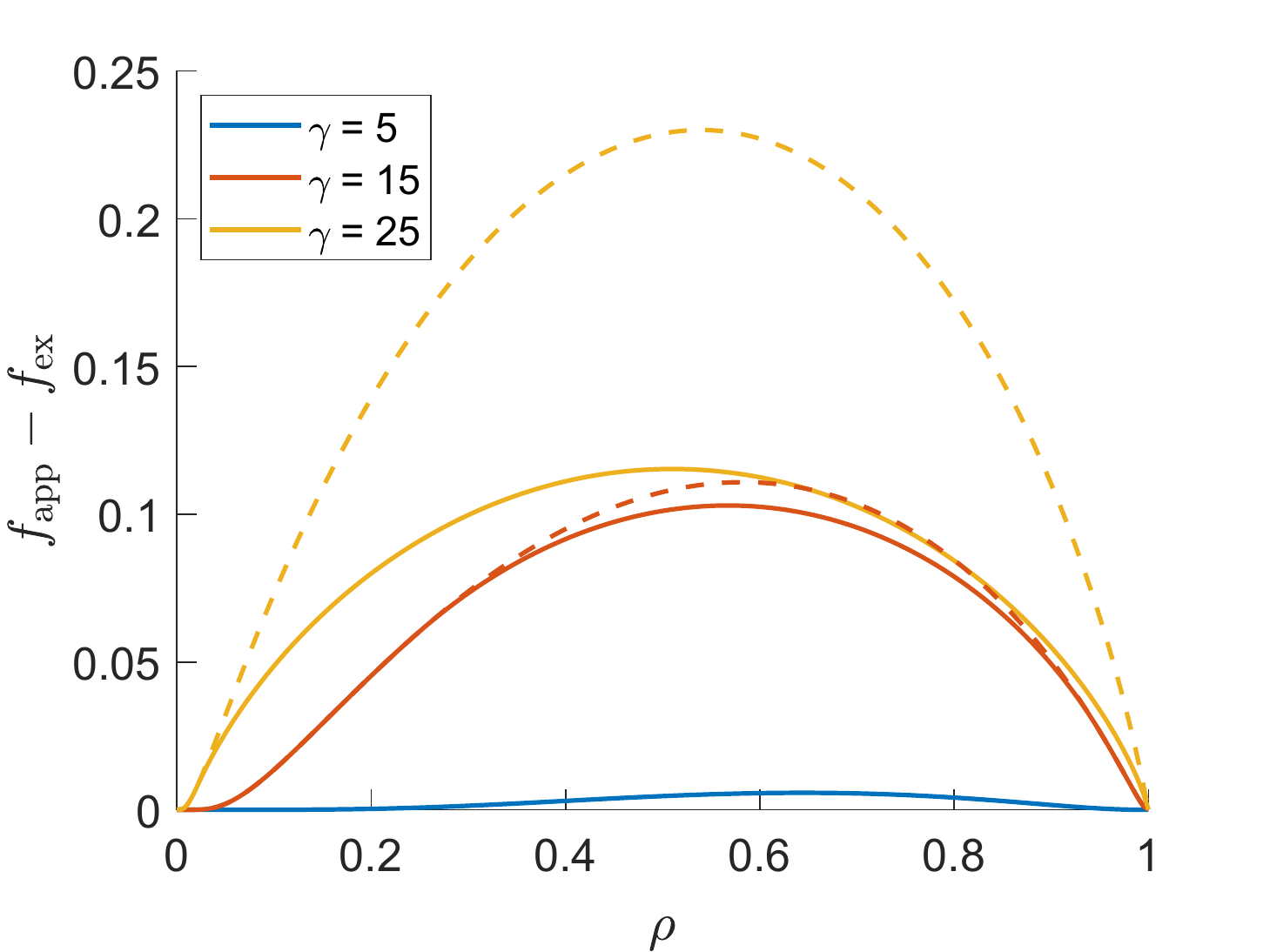}
\caption{\label{fig:N4} Dash lines represent the difference between the approximate free energy ($f_{\text{app}})$, given by Eq.~\eqref{eq:fN}, and the exact free energy ($f_{\text{ex}}$), given by Eq.~\eqref{eq:exactN4}, for the complete graph with $N=4$ nodes and for three values of the interaction parameter $\gamma$ (below, just above, and well above the pseudo-critical point). As the free energy has an unphysical concave region above the critical point, we also represent in solid lines the difference between the \emph{convex envelope} of the approximate free energy and the exact one, since this convex envelope is a better approximation to the real free energy.}
\end{figure}

In spite that the uniform free energy \eqref{eq:fN} predicts a critical point and a first-order phase transition for arbitrary $N$ (as low as $N=4$, see Fig.~\ref{fig:gammacN}), we know that this is not possible; in other words, this phase transition is not real. The reason because the free energy exhibits a concavity for some values of $\rho$ when $\gamma>\gamma_c(N)$ is that the equilibrium solution is not truly uniform for any value of the density---even though for some densities it is indistinguishable from a uniform one. The transition from the regions where the solution is almost uniform to those in which the structure is like those shown in Fig.~\ref{fig:histograms} (third row) is continuous---albeit probably abrupt. Thus, in the approximation we are using here, it shows up as a phase transition (i.e., the convex envelope of the free energy as a function of $\rho$ represents a good approximation of the true free energy of the system). In order to illustrate this point we compare our approximate free energy for $N=4$ with the exact one (obtained in Appendix~\ref{app:C}) in Fig.~\ref{fig:N4}.

\begin{figure}
\includegraphics[width=0.45\textwidth]{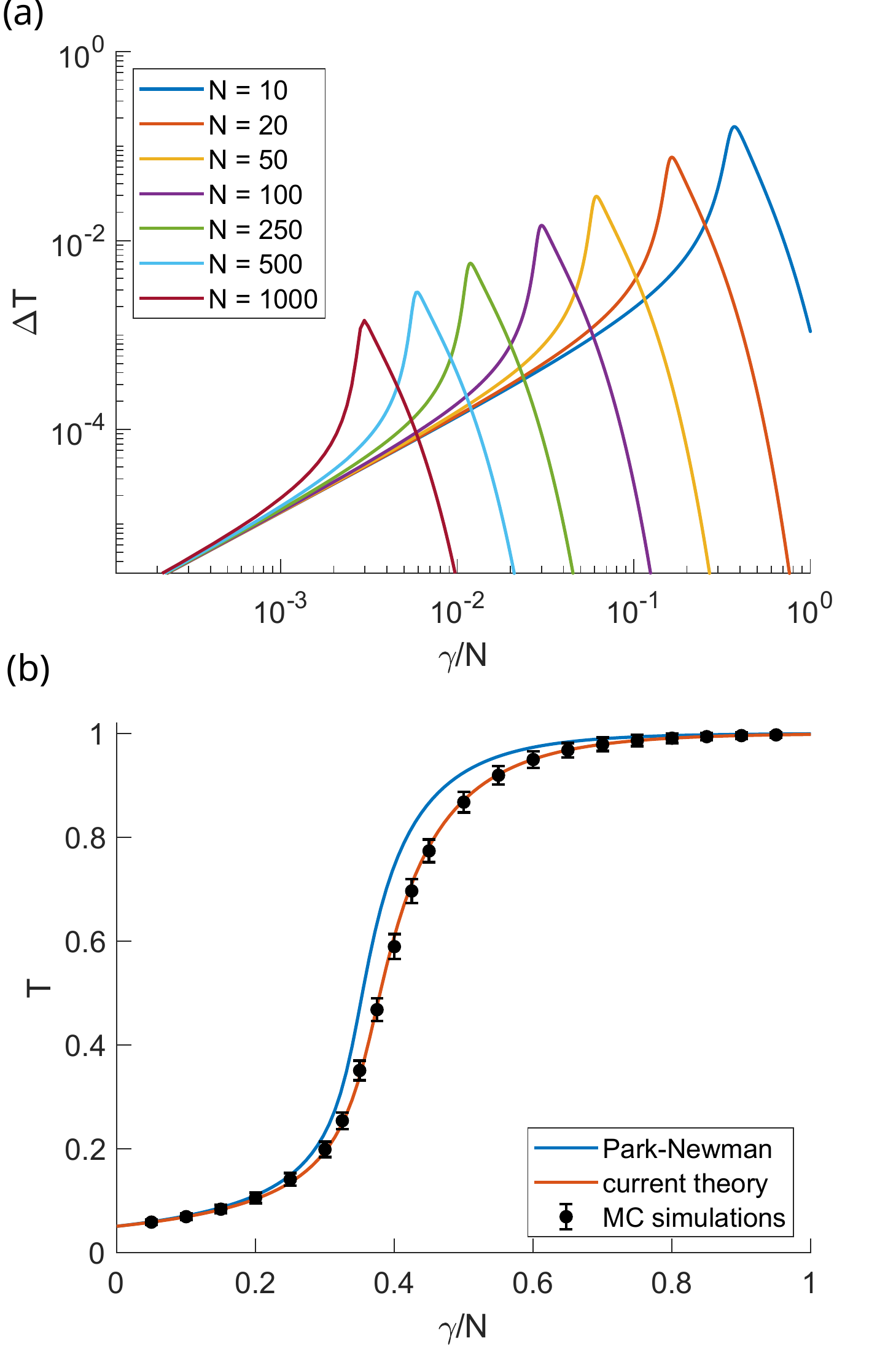}
\caption{\label{fig:difference} (a) Difference between the expected number of triangles $T$ as obtained from Park and Newman's mean-field calculations and from the current theory. (b) Expected number of triangles for $N=10$ according to both theories, along with Monte Carlo simulations. Error bars represent the standard deviation of the number of triangles along the simulations. In both panels $\phi=-0.53$.}
\end{figure}

\subsection{Comparison with Park and Newman's mean-field calculations}

A fair question to ask is how does the present theory compares with the mean-field calculations of Park and Newman \cite{park:2005}. In spirit, this theory is also mean-field-like, but clearly its construction follows a very different approach. On the other hand, because of the high dimensionality of this system one expects that in the thermodynamic limit it becomes exact \cite{park:2005}, so it would be desirable that, if not for all $N$, at least in this limit both theories coincide. Figure~3(b) of Ref.~\onlinecite{park:2005} shows the expected number of triangles $T$ (among other things) as a function of the interaction parameter $\gamma/N$, for $\phi=-0.53$ and $N=500$. We can obtain $T$ and $\phi$ as a function of $\rho$ and $\gamma$ through Eqs.~\eqref{eq:chempot} and \eqref{eq:triangleFMT}, respectively. From these two calculations, we can obtain parametrically the curve $T(\gamma)$ for fixed $\phi$ and for different values of $N$. The discrepancy between our results and those of Park and Newman is shown in Fig.~\ref{fig:difference}. Figure~\ref{fig:difference}(a) illustrates that the difference between the predictions of both theories decreases with system size---so that they both coincide in the thermodynamic limit. However, for very small networks their predictions differ significantly (e.g., for $N=10$, the discrepancy may be as high as $\sim 20\%$). Figure~\ref{fig:difference}(b) compares the predictions of both theories for $N=10$ along with Monte Carlo simulations performed using a Metropolis-Hastings algorithm \cite{snijders:2002}. Observable magnitudes are averaged over $10^6$ configurations of the Markov chain, taken one every $5\times 10^4$ steps. This figure highlights the higher accuracy of the current theory in calculating results for small networks.

Given that most real networks are large, the discussion of this section may seem as an academic issue of little practical relevance. Nevertheless, small networks of about 20--30 nodes are common e.g.~in social science, anthropology, or biology. Thus, one can find instances of these small networks in studies of different social organizations \cite{huitsing:2012,everett:2014,stadtfeld:2020,escribano:2021}, in bands of hunter-gatherers \cite{page:2017,migliano:2020}, or in groups of social animals \cite{kasper:2009,ilany:2013,escribano:2022}. For these studies, the improved accuracy provided by the present approach might not be negligible.

\bigskip

\section{\label{sec:nonhomogeneous}Non-homogeneous networks: Homophily}

Having an expression for the free energy of the non-homogeneous Strauss's model allows us to tackle other interesting cases. Particularly important is the case where there are different types of nodes in the network, with different interaction parameters. This case can model, e.g., homophily in a social network, where like nodes are more prone to form links or triangles than different nodes are \cite{mcpherson:2001}. A particular version of this model has already been used to study segregation on Strauss networks where triangles are both favored or disfavored \cite{avetisov:2018}.

\begin{figure}
\includegraphics[width=0.5\textwidth]{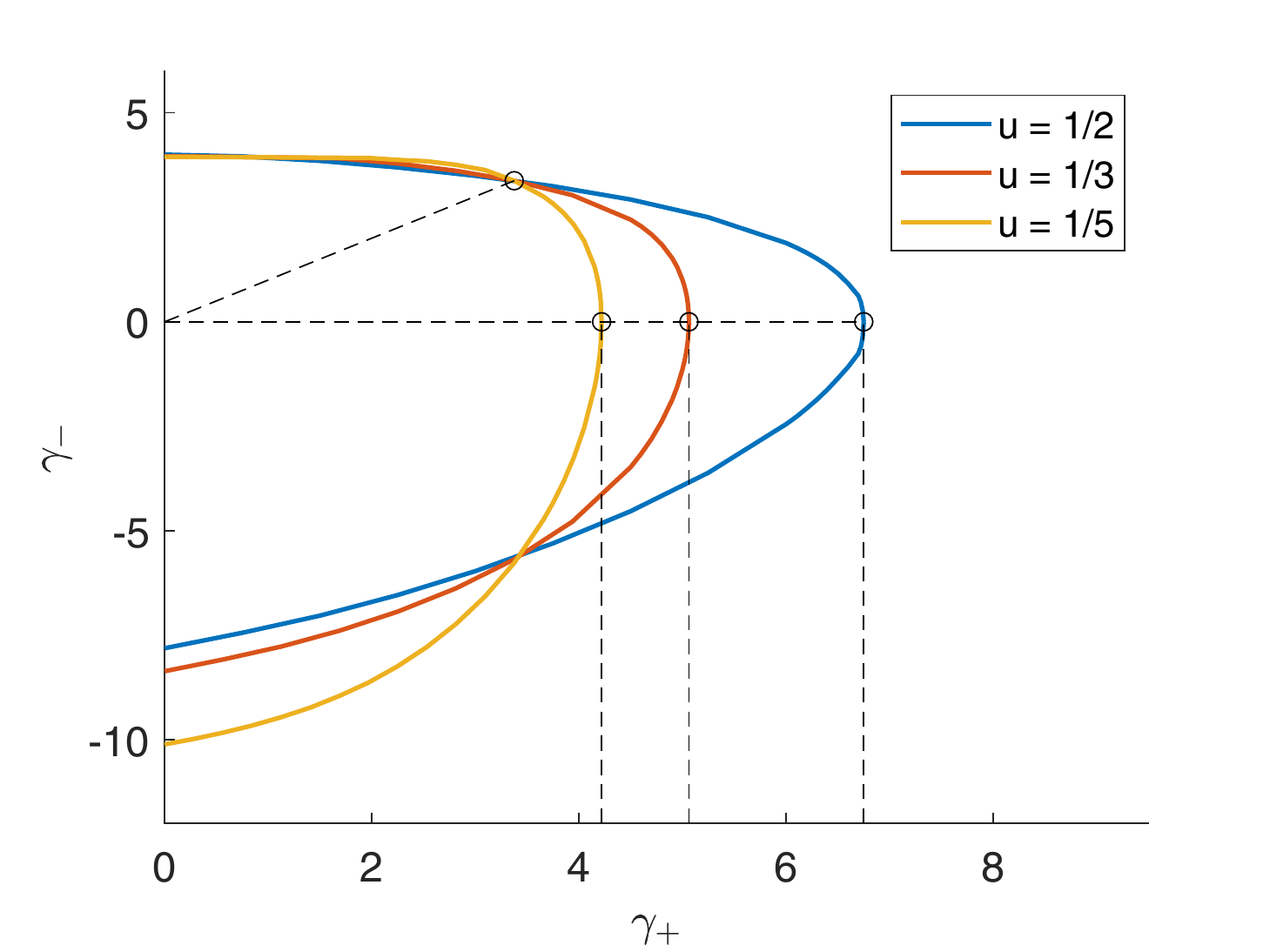}
\caption{\label{fig:criticalmix} Critical curves $\gamma_-$ vs.~$\gamma_+$ for different values of $0\le u\le 1/2$. The free energy of the non-homogeneous model is convex only for the points on the left of the curve. The curves reach their rightmost values of $\gamma_+$ (marked with circles and vertical dashed lines) for $\gamma_+=27/8(1-u)$, $\gamma_-=0$. The oblique dashed line is $\gamma_+=\gamma_-$. It meets all critical curves at one point (marked with a circle): $\gamma_+=\gamma_-=27/8$, the critical point of the homogeneous system.}
\end{figure}

Suppose we have two types of nodes in the network, A and B. Since the underlying graph is a complete graph, the actual location of these nodes is irrelevant, only how many of each type there are matters. So let us assume that there are $N_{\text{A}}$ of type A and $N_{\text{B}}=N-N_{\text{A}}$ of type B. Accordingly, $\binom{N_{\text{A}}}{2}$ links are homophilic of type AA, $\binom{N_{\text{B}}}{2}$ of type BB, and $N_{\text{A}}N_{\text{B}}$ are of mixed type. Likewise, there will be $\binom{N_{\text{A}}}{3}$ homophilic triangles of type AAA, $\binom{N_{\text{B}}}{3}$ of type BBB, $\binom{N_{\text{A}}}{2}N_{\text{B}}$ mixed triangles of type AAB, and $\binom{N_{\text{B}}}{2}N_{\text{A}}$ of type ABB. Hence, the free energy of the system can be obtained as
\begin{widetext}
\begin{equation}
    \begin{split}
        F=&\, \sum_{\text{X}=\text{A},\text{B}}
        \binom{N_{\text{X}}}{2}\big[\rho_{\text{XX}}\log\rho_{\text{XX}}
        +(1-\rho_{\text{XX}})\log(1-\rho_{\text{XX}})\big]
        +N_{\text{A}}N_{\text{B}}\big[\rho_{\text{AB}}\log\rho_{\text{AB}}
        +(1-\rho_{\text{AB}})\log(1-\rho_{\text{AB}})\big] \\
        &+\sum_{\text{X}=\text{A},\text{B}}
        \binom{N_{\text{X}}}{3}\left[3\rho_{\text{XX}}\log\left(1
        -\frac{\rho_{\text{XXX}}}{\rho_{\text{XX}}}\right)
        -2\log(1-\rho_{\text{XXX}})\right] \\
        &+\sum_{\text{X}=\text{A},\text{B}}\sum_{\text{Y}\ne\text{X}}
        \binom{N_{\text{X}}}{2}N_{\text{Y}}
        \left[\rho_{\text{XX}}\log\left(1-\frac{\rho_{\text{XXY}}}{\rho_{\text{XX}}}\right)
        +2\rho_{\text{XY}}\log\left(1-\frac{\rho_{\text{XXY}}}{\rho_{\text{XY}}}\right)
        -2\log(1-\rho_{\text{XXY}})\right],
    \end{split}
    \label{eq:Fmix}
\end{equation}
where $\rho_{\text{XY}}=\rho_{\text{YX}}$ is the density of links of type XY, and the densities associated to the triangles are the solutions of
\begin{equation}
    \zeta_{\text{XXY}}(\rho_{\text{XX}}-\rho_{\text{XXY}})
    (\rho_{\text{XY}}-\rho_{\text{XXY}})^2=\rho_{\text{XXY}}(1-\rho_{\text{XXY}})^2,
    \label{eq:rhoTmix}
\end{equation}

In order to reduce the number of parameters of the model we will henceforth assume that it is only homophily, and not the nature of the nodes, that determines interactions. This means that there are only two values of the interaction parameter instead of four, namely $\gamma_{\text{AAA}}=\gamma_{\text{BBB}}\equiv\gamma_+$, $\gamma_{\text{AAB}}=\gamma_{\text{BBA}}\equiv\gamma_-$. Furthermore, in the thermodynamic limit, the solution to \eqref{eq:rhoTmix} is
\begin{equation}
    \rho_{\text{XXY}}=\rho_{\text{XX}}\rho_{\text{XY}}^2\frac{\gamma_{\pm}}{N}
    +O\left(\frac{1}{N^2}\right),
\end{equation}
where the subindex of $\gamma_{\pm}$ depends on whether X$=$Y ($+$) or X$\ne$Y ($-$). In this same limit, and setting $N_{\text{A}}=uN$, $N_{\text{B}}=(1-u)N$, the free energy per link $f\equiv\binom{N}{2}^{-1}F$ turns out to be
\begin{equation}
    \begin{split}
        f=&\, u^2\big[\rho_{\text{AA}}\log\rho_{\text{AA}}
        +(1-\rho_{\text{AA}})\log(1-\rho_{\text{AA}})\big]
        +(1-u)^2\big[\rho_{\text{BB}}\log\rho_{\text{BB}}
        +(1-\rho_{\text{BB}})\log(1-\rho_{\text{BB}})\big] \\
        &+2u(1-u)\big[\rho_{\text{AB}}\log\rho_{\text{AB}}
        +(1-\rho_{\text{AB}})\log(1-\rho_{\text{AB}})\big]
        -\frac{\gamma_+}{3}\big[u^3\rho_{\text{AA}}^3+(1-u)^3\rho_{\text{BB}}^3\big] \\
        &-\gamma_-u(1-u)\rho_{\text{AB}}^2\big[u\rho_{\text{AA}}+(1-u)\rho_{\text{BB}}\big].
    \end{split}
    \label{eq:fmix2}
\end{equation}
The convexity of this function is linked to the positive definiteness of its Hessian matrix (entries are ordered as AA, BB, and AB)
\begin{equation}
    H=\begin{pmatrix}
    \displaystyle
    \frac{u^2}{\rho_{\text{AA}}(1-\rho_{\text{AA}})}-2\gamma_+u^3\rho_{\text{AA}} & 0 &
    -2\gamma_-u^2(1-u)\rho_{\text{AB}} \\[3mm]
    0 & \displaystyle
    \frac{(1-u)^2}{\rho_{\text{BB}}(1-\rho_{\text{BB}})}-2\gamma_+(1-u)^3\rho_{\text{BB}} &
    -2\gamma_-u(1-u)^2\rho_{\text{AB}} \\[3mm]
    -2\gamma_-u^2(1-u)\rho_{\text{AB}} & -2\gamma_-u(1-u)^2\rho_{\text{AB}} &
    \displaystyle
    \frac{2u(1-u)}{\rho_{\text{AB}}(1-\rho_{\text{AB}})}-2\gamma_-u(1-u)\bar\rho(u)
    \end{pmatrix},
\end{equation}
\end{widetext}
where we have introduced the shorthand notation $\bar\rho(u)\equiv u\rho_{\text{AA}}+(1-u)\rho_{\text{BB}}$. This translates into the positiveness of the first two diagonal elements plus $\det H>0$, in other words,
\begin{equation}
    \begin{split}
        &\chi_{\text{AA}}\equiv\frac{1}{\rho_{\text{AA}}(1-\rho_{\text{AA}})}-
        2\gamma_+u\rho_{\text{AA}}>0, \\
        &\chi_{\text{BB}}\equiv\frac{1}{\rho_{\text{BB}}(1-\rho_{\text{BB}})}-
        2\gamma_+(1-u)\rho_{\text{BB}}>0,
    \end{split}
    \label{eq:posdef1}
\end{equation}
and, removing trivial positive factors,
\begin{equation}
    \chi_{\text{AA}}\chi_{\text{BB}}\chi_{\text{AB}}
    -2u(1-u)\gamma_-^2\rho_{\text{AB}}^2(\chi_{\text{AA}}+\chi_{\text{BB}})>0,
    \label{eq:posdef2}
\end{equation}
with
\begin{equation}
    \chi_{\text{AB}}\equiv\frac{1}{\rho_{\text{AB}}(1-\rho_{\text{AB}})}
    -\gamma_-\bar\rho(u).
\end{equation}
Without loss of generality we may assume $0\le u\le 1/2$. With this assumption, inequalities \eqref{eq:posdef1} hold for any set of densities provided
\begin{equation}
    \gamma_+<\frac{27}{8(1-u)}.
\end{equation}
For any $\gamma_+$ satisfying this constraint, the critical value of $\gamma_-$ is obtained as the smallest value for which inequality \eqref{eq:posdef2} breaks down for some set of densities. This curve is represented in Fig.~\ref{fig:criticalmix} for several values of $u$.

Within the region enclosed by these curves, the function \eqref{eq:fmix2} correctly describes the system for given values of the two interaction parameters and the three densities. Outside this region, there are values of the densities for which the network is no longer stable as a uniform system; instead, it fractionates into coexisting subnetworks with different values of those densities.

A detailed analysis of the phase coexistence in the ternary mixture of links that describes this system goes beyond the scope of this work and will be dealt with in a subsequent study. The real purpose of this section is to illustrate how the functional we have derived can handle variations of the original system like this one.

\begin{figure}
\includegraphics[width=0.5\textwidth]{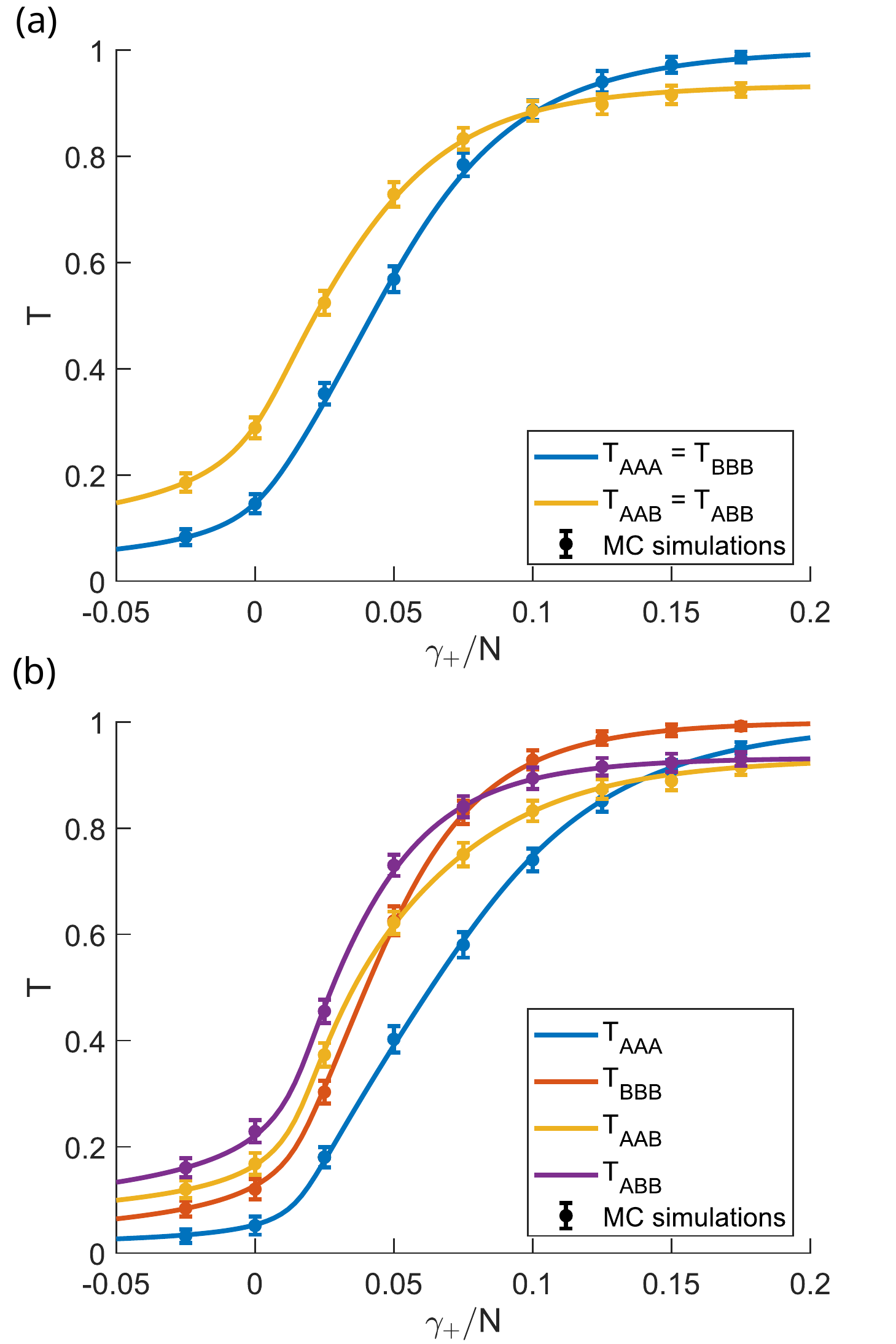}
\caption{\label{fig:mixsimulations} Fractions of the different kinds of triangles vs.~$\gamma_+/N$ for a system with $N=50$ nodes of two different types, A and B. The fraction of A nodes is $u=1/2$ (a) or $u=2/5$ (b). Solid lines are the curves obtained from  \eqref{eq:triangleFMT} and bullet points the Monte Carlo results. Error bars represent the standard deviation of the number of triangles along the simulations. In both panels $\gamma_-/N= 0.04$ and $\phi_{\text{AA}}=\phi_{\text{BB}}=\phi_{\text{AB}}=-0.25$.}
\end{figure}

In order to validate the expression of the free energy \eqref{eq:fmix2} we have performed Monte Carlo simulations (same method as before) to calculate the fractions of the different kinds of triangles, as functions of $\gamma_+$, for a system with $N=50$ nodes, for two values of $u=N_{\text{A}}/N$ ($1/2$ and $2/5$), and for fixed values of the other parameters ($\gamma_-/N= 0.04$ and $\phi_{\text{AA}}=\phi_{\text{BB}}=\phi_{\text{AB}}=-0.25$). Analytic expressions for those fractions of triangles are obtained from \eqref{eq:triangleFMT}. As in the case of uniform nodes, the agreement between theory and simulations (Fig.~\ref{fig:mixsimulations}) suggests that the expression of the free energy \eqref{eq:fmix2} might be exact in the thermodynamic limit.

\section{\label{sec:discussion}Discussion and conclusions}

In this paper, we have solved approximately Strauss's model of transitive networks using a technique specific of the statistical physics of lattice gases---density-functional theory. The solution we have found is more accurate than a standard mean-field approximation for small systems, but coincides with it (and probably with the exact solution) in the thermodynamic limit of infinitely many nodes. The model exhibits a first-order phase transition for triangle interactions above a critical threshold $\gamma_c$. For $\gamma>\gamma_c$, upon increasing the probability that links are created the system crosses a region where two solutions are possible---one with low and one with high fraction of links (density). Because of this fact, this model has been deemed unsuitable to produce networks with intermediate fractions of links.

The density-functional formalism that we have employed reveals that the canonical ensemble (constant density and ``temperature'', i.e., triangle interaction) is the natural description for this system if we want to access this ``forbidden'' intermediate states. In this ensemble, the system behaves as a fluid undergoing a condensation transition. The two (low and high density) phases are akin to a gas and a liquid, and at those intermediate densities both phases coexist in chemical and mechanical equilibrium. A histogram of links belonging to a given number of triangles shows that links in the graph form two separated groups, each associated to one of these two phases. Hence, graphs within this coexisting region have a different structure to those out of it.

Under this interpretation, the problem of generating graphs with this model in the ``inaccessible'' coexisting region amounts to performing Monte Carlo simulations using Kawasaki dynamics, which keeps the number of links constant. The idea of accessing these intermediate states by controlling for an extensive parameter has already been suggested \cite{newman:2009,miller:2009}---although the preferred control variable has always been the number of triangles.

Whether the graphs produced by Strauss's model are a suitable model for some real networks is still an open question. It is true that the peculiar structure of these graphs has never been observed so far, but it is also true that the existence of the phase transition in Strauss's model (and its extensions) is an unavoidable consequence of the specific interaction among its links. We still do not know whether a real system with that particular kind of interactions can be observed or even devised.

The density-functional formalism that we have developed here can further be applied to systems in which the interaction constants are link- or triangle-dependent. This way we can study systems in which nodes have different types and interactions depending on the type of the nodes involved. Homophily is one of the situations that can be so described. The analysis of the simplest example of homophilic interactions shows that homophily favors the stability of uniform networks (networks with uniform density) by increasing the value of the critical point. The predictions for this case have been validated with Monte Carlo simulations, which---as in the case of uniform networks---seem to suggest that the free energy here obtained might be exact in the thermodynamic limit. Further studies are needed to achieve a full characterization of the complex phase behavior of a system like this.

Finally, we would like to emphasize that the main contribution of this paper is to provide a formalism that can be extended to tackle other ERG models---its advantage with respect to more standard mean-field approaches being that it provides a systematic procedure to deal with them. Hopefully, it may become a useful tool to analyze this class of network models.

\begin{acknowledgments}
This research is part of project BASIC (PGC2018-098186-B-I00) funded by MCIN / AEI / 10.13039 / 501100011033 and by ``ERDF A way of making Europe''.
\end{acknowledgments}

\appendix

\section{\label{app:A}Free energy of a single triangle}

For a triangle,
\begin{equation}
    \Xi_3=(1+z_{ij})(1+z_{jk})(1+z_{ki})+\zeta_{ijk}z_{ij}z_{jk}z_{ki},
    \label{eq:Xi3z}
\end{equation}
hence
\begin{equation}
    \begin{split}
        \rho_{ij} &=\frac{z_{ij}(1+z_{jk})(1+z_{ki})+\zeta_{ijk}z_{ij}z_{jk}z_{ki}}{\Xi_3}, \\
        \rho_{jk} &=\frac{z_{jk}(1+z_{ij})(1+z_{ki})+\zeta_{ijk}z_{ij}z_{jk}z_{ki}}{\Xi_3}, \\
        \rho_{ki} &=\frac{z_{ki}(1+z_{ij})(1+z_{jk})+\zeta_{ijk}z_{ij}z_{jk}z_{ki}}{\Xi_3}.
    \end{split}
    \label{eq:rhosz}
\end{equation}
It will prove convenient to introduce
\begin{equation}
    \rho_{ijk}\equiv\frac{\zeta_{ijk}z_{ij}z_{jk}z_{ki}}{\Xi_3}.
    \label{eq:rhoabc}
\end{equation}
Now, dividing \eqref{eq:Xi3z} by $\Xi_3$ we get
\begin{equation}
    \frac{(1+z_{ij})(1+z_{jk})(1+z_{ki})}{\Xi_3}=1-\rho_{ijk}.
    \label{eq:intermediate}
\end{equation}
On the other hand, \eqref{eq:rhosz} can be rewritten as
\begin{equation}
    \begin{split}
        \rho_{ij}-\rho_{ijk} &=\frac{z_{ij}(1+z_{jk})(1+z_{ki})}{\Xi_3}, \\
        \rho_{jk}-\rho_{ijk} &=\frac{z_{jk}(1+z_{ij})(1+z_{ki})}{\Xi_3}, \\
        \rho_{ki}-\rho_{ijk} &=\frac{z_{ki}(1+z_{ij})(1+z_{jk})}{\Xi_3}.
    \end{split}
    \label{eq:rhoszs}
\end{equation}
Multiplying them out and using \eqref{eq:rhoabc} and \eqref{eq:intermediate} leads to Eq.~\eqref{eq:cubic}. Also, using \eqref{eq:intermediate} in \eqref{eq:rhoszs} we obtain
\begin{equation}
    \begin{split}
        \rho_{ij}-\rho_{ijk} &=\frac{z_{ij}}{1+z_{ij}}(1-\rho_{ijk}), \\
        \rho_{jk}-\rho_{ijk} &=\frac{z_{jk}}{1+z_{jk}}(1-\rho_{ijk}), \\
        \rho_{ki}-\rho_{ijk} &=\frac{z_{ki}}{1+z_{ki}}(1-\rho_{ijk}),
    \end{split}
\end{equation}
whose solutions are
\begin{equation*}
    z_{ij} =\frac{\rho_{ij}-\rho_{ijk}}{1-\rho_{ij}}, \quad
    z_{jk} =\frac{\rho_{jk}-\rho_{ijk}}{1-\rho_{jk}}, \quad
    z_{ki} =\frac{\rho_{ki}-\rho_{ijk}}{1-\rho_{ki}}.
\end{equation*}
Substituting these expressions in \eqref{eq:rhoabc} and using \eqref{eq:cubic} we obtain
\begin{equation}
    \Xi_3=\frac{(1-\rho_{ijk})^2}{(1-\rho_{ij})(1-\rho_{jk})(1-\rho_{ki})}.
\end{equation}
Thus $\Phi_3=\rho_{ij}\log z_{ij}+\rho_{jk}\log z_{jk}+\rho_{ki}\log z_{ki}-\log\Xi_3$ becomes \eqref{eq:Phi3}.

\section{\label{app:B}Probability of forming a triangle}

The probability that nodes $i,j,k$ form a triangle can be obtained from the grand potential as
\begin{equation}
T_{ijk}\equiv\langle\tau_{ij}\tau_{jk}\tau_{ki}\rangle
=-N\frac{\partial\Omega}{\partial\gamma_{ijk}}.
\end{equation}
But inverting the Legendre transform \eqref{eq:Legendre},
\begin{equation}
    \Omega=F-\sum_{\{ij\}}\phi_{ij}\rho_{ij},
\end{equation}
where $\rho_{ij}$ depends $\bm\phi$ and $\bm\gamma$ through \eqref{eq:chempot}. Thus,
\begin{equation}
T_{ijk}=\sum_{\{lm\}}\phi_{lm}
N\frac{\partial\rho_{lm}}{\partial\gamma_{ijk}}
-N\frac{\partial F}{\partial\gamma_{ijk}}.
\label{eq:triangleijk}
\end{equation}
Now,
\begin{equation*}
\frac{\partial F}{\partial\gamma_{ijk}}=\sum_{\{lm\}}
\frac{\partial F}{\partial\rho_{lm}}\frac{\partial\rho_{lm}}{\partial\gamma_{ijk}}+
\left(\frac{\partial F}{\partial\gamma_{ijk}}\right)_{\bm\rho},
\end{equation*}
where the last partial derivative is taken at constant $\bm\rho$. Thus, substituting into \eqref{eq:triangleijk} and using \eqref{eq:chempot} we obtain
\begin{equation}
T_{ijk}=-N\left(\frac{\partial F}{\partial\gamma_{ijk}}\right)_{\bm\rho}
=-(1+\zeta_{ijk})\left(\frac{\partial F}{\partial\zeta_{ijk}}\right)_{\bm\rho}.
\end{equation}
Notice that the free energy depends on $\zeta_{ijk}$ only through $\rho_{ijk}$ via Eq.~\eqref{eq:cubic}, therefore
\begin{align*}
T_{ijk}=&\, -(1+\zeta_{ijk})\left(\frac{\partial F}{\partial\rho_{ijk}}\right)_{\bm\rho}
\left(\frac{\partial\rho_{ijk}}{\partial\zeta_{ijk}}\right)_{\bm\rho} \\
=&\, (1+\zeta_{ijk})\rho_{ijk}\left[\frac{1-3\rho_{ijk}}{\rho_{ijk}(1-\rho_{ijk})}
+\frac{1}{\rho_{ij}-\rho_{ijk}} \right. \\
+&\left. \frac{1}{\rho_{jk}-\rho_{ijk}}+\frac{1}{\rho_{ki}-\rho_{ijk}}\right]
\left(\frac{\partial\rho_{ijk}}{\partial\zeta_{ijk}}\right)_{\bm\rho}.
\end{align*}
On the other hand, taking logarithms of \eqref{eq:cubic} and differentiating with respect to $\zeta_{ijk}$ at constant $\bm\rho$ we get
\begin{align*}
\frac{1}{\zeta_{ijk}}=&\,\left[\frac{1-3\rho_{ijk}}{\rho_{ijk}(1-\rho_{ijk})}
+\frac{1}{\rho_{ij}-\rho_{ijk}} \right. \\
+&\left. \frac{1}{\rho_{jk}-\rho_{ijk}}+\frac{1}{\rho_{ki}-\rho_{ijk}}\right]
\left(\frac{\partial\rho_{ijk}}{\partial\zeta_{ijk}}\right)_{\bm\rho},
\end{align*}
which finally leads to \eqref{eq:triangleFMT}.

\section{Strauss's model for small networks}
\label{app:C}

Let $\bm\tau$ denote a vector of components $\tau_{\nu}$, where $\nu$ is a subset of $2$ elements of $\{1,2,\dots,N\}$, and $W\equiv\{0,1\}^N$. The grand partition function is defined as
\begin{equation}
    \begin{split}
        \Xi &=\sum_{\boldsymbol{\sigma}\in W}\exp\left(\phi\sum_{i<j}^N\tau_{ij}+
        \frac{\gamma}{N}\sum_{i<j<k}^N\tau_{ij}\tau_{jk}\tau_{ik}\right) \\
        &=\sum_{L=0}^{\binom{N}{2}}\sum_{T=0}^{\binom{N}{3}}Q(L,T)e^{\phi L+\gamma T/N},
    \end{split}
\end{equation}
where $Q(L,T)$ is the number of configurations $\bm\tau$ with $L$ links and $T$ triangles.

Let us set $N=4$ and compute the values of $Q(L,T)$. Clearly, $Q(L,0)=\binom{6}{L}$ for $L=0,1,2$, but $Q(L,T)=0$ otherwise. Furthermore, $Q(5,2)=6$, $Q(6,4)=1$, and  $Q(5,T)=Q(6,T)=0$ otherwise. As for $L=3,4$, there are configurations with either $T=0$ or $T=1$. Thus, $Q(3,1)=4$ and $Q(3,0)=\binom{6}{3}-4=16$. On the other hand, $Q(4,0)=3$ and $Q(4,1)=\binom{6}{4}-3=12$. Accordingly, if we denote $x\equiv e^{\phi}$, $y\equiv e^{\gamma/4}$,
\begin{equation}
    \begin{split}
        \Xi=&\, 1+6x+15x^2+16x^3+3x^4+4x^3(1+3x)y \\
        &+6x^5y^2+x^6y^4.
    \end{split}
\end{equation}
The density can be obtained as
\begin{equation}
    \begin{split}
        \rho &=\frac{1}{6}\frac{\partial}{\partial\phi}\log\Xi
        =\frac{x}{6}\frac{\partial}{\partial x}\log\Xi, 
    \end{split}
\end{equation}
and from that,
\begin{equation}
    f=\frac{F}{6}=\rho\log x-\frac{1}{6}\log\Xi.
    \label{eq:exactN4}
\end{equation}
Thus, fixing the interaction $\gamma$ (i.e., fixing $y$) we can obtain parametrically, as $0<x<\infty$, the curve $f(\rho,\gamma)$.

\bibliography{Strauss}

\end{document}